\documentclass[runinaddress,showpacs,twocolumn,prb]{revtex4}
\usepackage{amssymb}
\usepackage{amsmath}
\usepackage{graphicx}

\begin{document}

\title{Deduction of Pure Spin Current from Spin Linear and Circular
Photogalvanic Effect in Semiconductor Quantum Wells}
\date{\today}
\author{Bin Zhou$^{1,2}$ and Shun-Qing Shen$^{1}$}
\affiliation{$^{1}$Department of Physics, and Center for Theoretical and Computational
Physics, The University of Hong Kong, Pokfulam Road, Hong Kong, China\\
$^{2}$Department of Physics, Hubei University, Wuhan 430062, China}

\begin{abstract}
We study the spin photogalvanic effect in two-dimensional electron system
with structure inversion asymmetry by means of the solution of semiconductor
optical Bloch equations. It is shown that a linearly polarized light may
inject a pure spin current in spin-splitting conduction bands due to Rashba
spin-orbit coupling, while a circularly polarized light may inject
spin-dependent photocurrent. We establish an explicit relation between the
photocurrent by oblique incidence of a circularly polarized light and the
pure spin current by normal incidence of a linearly polarized light such
that we can deduce the amplitude of spin current from the measured spin
photocurrent experimentally. This method may provide a source of spin
current to study spin transport in semiconductors quantitatively.
\end{abstract}

\pacs{72.25.Fe, 72.40.+w}
\maketitle

\section{Introduction}

Spin-coherent transport of conduction electrons in semiconductor
heterostructures is currently an emerging subject due to its possible
application in a new generation of electronic devices.\cite{Awschalom02Book}
There have been considerable efforts to achieve spin-polarized current or
pure spin current (PSC) in semiconductors. Optical injection of spin current
is based largely on the fact that the spin-polarized carriers in conduction
band can be injected in semiconductors via absorption of the polarized
light. In the case of semiconductors, if the photon energy is higher than
the characteristic energy gap, such as that of the conduction and valence
bands of electrons, or of intersubband, electrons are pumped into the
conduction band from the valence band or conduction subband. When the system
breaks the inversion symmetry, the single-photon absorption may generate
spin current or spin polarized current. The circular photogalvanic effect
(CPGE), which is based on converting the helicity of light into an electric
current by irradiation of circularly polarized light, was studied
extensively. \cite%
{Ivchenko78SPJ,Belinicher78PLA,Averkiev83SP,Hagele98APL,Kikkawa99nature,Ganichev00APL,Ganichev01PRL,Ganichev03JPCM,Yang06PRL}
Conventional CPGE focus only on the charge, not spin aspect of electronic
transport in semiconductors. It was first realized that quantum interference
of one- and two-photon excitation of unbiased semiconductors may yield
ballistic spin-polarized current, which was observed by two groups.\cite%
{Hubner03PRL,Stevens03PRL,Najmaie05PRL} Recently it was proposed that
one-photon absorption of linearly polarized light should produce PSC in the
noncentrosymmetric semiconductors.\cite%
{Bhat05PRL,Zhao05PRB,Tarasenko05JETP,Tarasenko06xxx} On the other hand,
experimental detection of PSC has been realized by measuring spin
accumulation near the boundary of samples \cite%
{Kato04Science,Wunderlich05PRL} and electric current induced by PSC in a
crossbar system.\cite{Tinkham06Nature,Cui06xxx,Li06APL}

Structure inversion asymmetry (SIA) in semiconductor heterojunctions may
lead to spin splitting of the conduction band in the momentum $\mathbf{k}$%
-space, and induce the spin-orbit coupling. This type of the systems may be
one of good candidates to implement spin-based electronic devices and has
attracted more and more attentions. In this paper, we investigate how to
deduce PSC in the semiconductor quantum wells (QWs) by irradiation of
linearly and circularly polarized lights. By using the solution of the
semiconductor optical Bloch equations, we establish an explicit relation
between spin photocurrent by oblique incidence of a circularly polarized
light and PSC with in-plane spin polarization by normal incidence of a
linearly polarized light with the same frequency and intensity. Since the
photocurrent can be measured experimentally, we can deduce PSC from measured
photocurrent based on several material specific parameters. This method can
provide an efficient source for generating PSC quantitatively, and has
potential applications in semiconductor spintronics.

\section{ Model and general formalism}

We consider a QW of zinc-blende-type semiconductors with SIA. The conduction
electrons can be modeled as
\begin{equation}
H_{c}=\frac{\hbar ^{2}k^{2}}{2m^{\ast }}-\lambda \hbar \left( k_{x}\sigma
_{y}-k_{y}\sigma _{x}\right) ,
\end{equation}%
where $\mathbf{\sigma }$ are the Pauli matrices, $\lambda $ is the strength
of Rashba spin-orbit coupling, and $m^{\ast }$ is the effective mass of
conduction electron. The valence bands near the $\Gamma $ point are
described approximately by the Luttinger Hamiltonian for spin $S=3/2$ holes,
\begin{equation}
H_{L}=-\frac{\hbar ^{2}}{2m}\left[ \left( \gamma _{1}+\frac{5}{2}\gamma
_{2}\right) k^{2}-2\gamma _{2}\left( \mathbf{k}\cdot \mathbf{S}\right) ^{2}%
\right] ,
\end{equation}%
where $\gamma _{1}$, $\gamma _{2}$ are two Kohn-Luttinger parameters, $m$ is
the free electron mass and $\mathbf{S}$ represents three $4\times 4$ spin $%
3/2$ matrices. For a bulk system, both heavy- and light-hole bands are
degenerate at the $\Gamma $ point. In a QW with thickness $L$, while $k_{x}$
and $k_{y}$ are good quantum numbers, the confinement along the $z$-axis is
approximately realized by taking $\left\langle k_{z}\right\rangle =0$, and $%
\left\langle k_{z}^{2}\right\rangle \simeq \left( \pi /L\right) ^{2}$ for
the lowest energy band. In the case of $k^{2}=k_{x}^{2}+k_{y}^{2}\ll
\left\langle k_{z}^{2}\right\rangle $, the energy spectrum of the first
doubly degenerated heavy-hole band is reduced approximately into $%
E^{HH}\simeq -\hbar ^{2}k^{2}/\left( 2m_{HH}\right) -\varepsilon $, with the
effective mass $m_{HH}=m/\left( \gamma _{1}+\gamma _{2}\right) $, and $%
\varepsilon =\hbar ^{2}\left\langle k_{z}^{2}\right\rangle \left( \gamma
_{1}-2\gamma _{2}\right) /(2m)$. Finite thickness of the QW makes the band
structure into a sequence of quasi-two-dimensional (2D) subbands with $%
\left\langle k_{z}^{2}\right\rangle \simeq \left( n\pi /L\right) ^{2}$ ($n$
is a non-zero integer), which can be calculated numerically.\cite%
{Winkler2003} Of course, for the precise calculations, we need to take into
account the band structure of the whole $\mathbf{k}$-space. In the present
paper, we first consider this simplified 2D model and then present numerical
results by taking into account the finite thickness effect of band structure
near the $\Gamma $ point.

\begin{figure}[tbp]
\includegraphics[width=7cm]{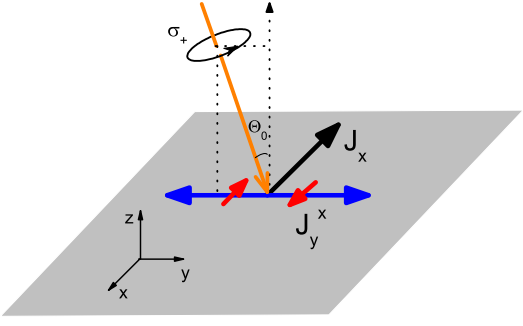}
\caption{{}The sketch of a right-handed circularly polarized ($\protect%
\sigma _{+}$) light irradiating on the surface of a semiconductor QW in the (%
$yz$) plane with incidence angle $\Theta _{0}$. In this case, the
photocurrent $J_{x}$ is injected perpendicularly to the incident plane of
the light, and PSC $J_{y}^{x}$ is also injected. Orange arrow denotes the
direction of light propagation; thick black arrow denotes photocurrent; blue
arrows denote PSC with in-plane spin polarization represented by red arrows.}
\end{figure}

Now we come to study the irradiation of a polarized light on the system with
incidence angle $\Theta _{0}$ in the plane ($yz$) as shown in Figure 1. The
pump pulse is of the form
\begin{equation}
\mathbf{E}\left( t\right) =\mathbf{E}_{\omega }e^{-i(\omega t-k\cos \Theta
_{0}z+k\sin \Theta _{0}y)}+c.c.,
\end{equation}%
where $\omega $ is the frequency of the light. By treating the field
perturbatively, and assuming fast interband dephasing, the semiconductor
optical Bloch equations give the single particle density matrix in
conduction bands due to optical irradiation \cite%
{Bhat05PRL,HaugBook,Rossler02pss,Golub03PRB}%
\begin{eqnarray}
\rho _{cc^{\prime }}\left( \mathbf{k}\right) &=&\frac{\pi e^{2}}{\hbar ^{2}}%
\sum_{v}\frac{\mathbf{E}_{\omega }\cdot \mathbf{v}_{cv}}{\omega _{cv}}\frac{%
\mathbf{E}_{\omega }^{\ast }\cdot \mathbf{v}_{vc^{\prime }}}{\omega
_{c^{\prime }v}}  \notag \\
&&\times \left[ \delta \left( \omega -\omega _{cv}\right) +\delta \left(
\omega -\omega _{c^{\prime }v}\right) \right] \tau _{e},
\end{eqnarray}%
where the subscripts $c$ and $v$ refer to conduction and valence bands, $%
\mathbf{v}_{cv}\left( \mathbf{k}\right) =$ $\left\langle c\mathbf{k}%
\right\vert \mathbf{v}\left\vert v\mathbf{k}\right\rangle $ is the interband
matrix element of the velocity operator, $\tau _{e}$ is the momentum
relaxation time as a result of all various interactions, $\hbar \omega
_{c\left( c^{\prime }\right) v}=\hbar ^{2}k^{2}/(2\mu )\pm \lambda \hbar
k+\Delta _{0}$ (with $\Delta _{0}$ being the band gap and the\ reduced mass $%
\mu =m^{\ast }m_{HH}/(m^{\ast }+m_{HH})$) for the simplified 2D model. Using
this solution, a physical observable $\mathbf{O}$ in conduction bands can be
calculated by%
\begin{equation}
\mathbf{O}=\sum_{c,c^{\prime },\mathbf{k}}\left\langle c^{\prime }\mathbf{k}%
\right\vert \hat{O}\left\vert c\mathbf{k}\right\rangle \rho _{cc^{\prime
}}\left( \mathbf{k}\right) ,  \label{formula}
\end{equation}%
where $\hat{O}$ is the corresponding operator. In the following, spin
current operator $\hat{J}_{i}^{j}$ is defined conventionally as $\hat{J}%
_{i}^{j}=\frac{\hbar }{4}\left\{ v_{i},\sigma _{j}\right\} $.

\section{Spin circular photogalvanic effect (SCPGE)}

Spin photocurrent in the CPGE was studied extensively. Here we focus on spin
aspect of the CPGE. Consider oblique incidence of a circularly polarized
light onto the system. In this case a spin photocurrent can be circulated to
be perpendicular to the incident plane of the light. When the light enters
into the sample, due to the refraction effect, the light becomes $%
E_{x}=E_{0}t_{s}\cos \varphi $, $E_{y}=iE_{0}t_{p}\sin \varphi \cos \Theta $%
, and $E_{z}=iE_{0}t_{p}\sin \varphi \sin \Theta $, where $E_{0}$ is the
electric field amplitude in vacuum, $\Theta $ is the angle of refraction
defined by $\sin \Theta =\sin \Theta _{0}/n$ ( $n$ is the index of
refraction), $t_{s}=2\cos \Theta _{0}/(\cos \Theta _{0}+n\cos \Theta )$ and $%
t_{p}=2\cos \Theta _{0}/\left( n\cos \Theta _{0}+\cos \Theta \right) $ are
transmission coefficients after Fresnel's formula for linear $s$ and $p$
polarizations. \cite{Ivchenko97Book} The helicity of the incident light is $%
P_{\text{circ}}=\left( I_{\sigma _{+}}-I_{\sigma _{-}}\right) /\left(
I_{\sigma _{+}}+I_{\sigma _{-}}\right) =\sin 2\varphi $, where $I_{\sigma
_{+}}$ and $I_{\sigma _{-}}$ are intensities of right- ($\sigma _{+}$) and
left-handed ($\sigma _{-}$) polarized radiations. $P_{\text{circ}}=\pm 1$
denotes right and left circularly polarized light, respectively. In this way
the photocurrent can be calculated explicitly.\cite{Golub03PRB} The hole
current induced in the valence bands is neglected because the effective mass
of holes is typically much greater than that of electrons, and the kinetic
energy and speed of holes are much less than those of the electrons.\cite%
{Hache98IEEE} In the oblique incidence of a circularly polarized light in ($%
yz$) plane, the formula (\ref{formula}) gives the photocurrent $J_{y}=0$ and
\begin{equation}
J_{x}=-\frac{2\lambda \mu ^{3}\Omega }{3\hbar ^{5}\left\langle
k_{z}^{2}\right\rangle m^{\ast }}t_{s}t_{p}a_{0}^{2}e^{3}E_{0}^{2}\tau
_{e}P_{\text{circ}}\sin \Theta ,  \label{current}
\end{equation}%
where $\Omega =\lambda ^{2}\mu +\hbar \omega -\Delta _{0}$, and $a_{0}=\sqrt{%
6}\left\langle 0,0\right\vert x\left\vert 1,-1\right\rangle $ which is a
parameter determined by experiment. It is clear that the photocurrent
circulates only in the case of the circularly polarized light ($P_{\text{circ%
}}\neq 0$), and vanishes in the case of linearly polarized light ($P_{\text{%
circ}}=0$). Besides the photocurrent in CPEG, a PSC with $x$-component spin
polarization perpendicular to the direction of photocurrent also circulates,%
\begin{equation}
J_{y}^{x}=(I_{+}^{C}t_{p}^{2}\cos ^{2}\Theta \sin ^{2}\varphi
+I_{-}^{C}t_{s}^{2}\cos ^{2}\varphi )\hbar a_{0}^{2}e^{2}E_{0}^{2}\tau _{e},
\label{JyxCO}
\end{equation}%
where
\begin{equation}
I_{\pm }^{C}=\frac{\lambda \mu }{6\hbar ^{5}\left\langle
k_{z}^{2}\right\rangle m^{\ast }}\left[ \hbar ^{2}\left\langle
k_{z}^{2}\right\rangle \left( m^{\ast }-\mu \right) \pm \mu ^{2}\Omega %
\right] .
\end{equation}%
The spin current even survives even in the normal incidence while the
photocurrent vanishes. It is sketched in Figure 1 that the photocurrent $%
J_{x}$ and PSC $J_{y}^{x}$ are induced by a right-handed circularly
polarized light irradiating on the surface of a semiconductor QW in the ($yz$%
) plane with incidence angle $\Theta _{0}$.

Absorption of a circularly polarized light in semiconductors induces $z$%
-component spin polarization $S^{z}$ due to the conservation of angular
momentum. The light-induced non-zero $S^{z}$ will lead to an orientational
distribution of PSC with the $z$-component polarization
\begin{equation}
J_{r}^{z}\left( \theta \right) =-\frac{\mu ^{2}\Omega }{6\hbar ^{3}m^{\ast
}\pi \delta }t_{s}t_{p}\hbar a_{0}^{2}e^{2}E_{0}^{2}\tau _{e}P_{\text{circ}%
}\cos \Theta ,  \label{Jtz-CO}
\end{equation}%
with $\theta =\arctan \left( k_{x}/k_{y}\right) $, $\delta =\sqrt{\mu \left(
\lambda ^{2}\mu +2\hbar \omega -2\Delta _{0}\right) }$, and the subscript $r$
denoting the radial direction in polar coordinates. However, one has $%
J_{x(y)}^{z}\left( \mathbf{k}\right) =J_{x(y)}^{z}\left( -\mathbf{k}\right) $
such that the total spin current with $z$-component polarization vanishes.
With the geometric constraint of the sample, a PSC of $z$-component
polarization can circulate and may be used to implement the reciprocal spin
Hall effect.\cite{Hankiewicz05PRB} In the case of normally incident $\sigma
^{+}$ polarized light, the orientational distributions of radial spin
current with tangent direction polarization (i.e., $J_{r}^{\theta }\left(
\theta \right) $) and tangent spin current with radial direction
polarization (i.e., $J_{\theta }^{r}\left( \theta \right) $) are given by%
\begin{equation}
J_{r}^{\theta }\left( \theta \right) =\frac{\lambda \mu \left( 2\mu -m^{\ast
}\right) }{12\hbar ^{3}m^{\ast }\pi }t_{0}^{2}\hbar
a_{0}^{2}e^{2}E_{0}^{2}\tau _{e},  \label{Jrt-CN}
\end{equation}%
\begin{equation}
J_{\theta }^{r}\left( \theta \right) =\frac{\lambda \mu }{12\hbar ^{3}\pi }%
t_{0}^{2}\hbar a_{0}^{2}e^{2}E_{0}^{2}\tau _{e},  \label{Jtr-CN}
\end{equation}%
where the sub- and super-script $\theta $ denotes the tangent direction in
polar coordinates. The total contribution of the orientational distributions
of spin current leads to a non-vanishing spin current with in-plane spin
polarization as shown in Eq. (\ref{JyxCO}).

For the sake of clarity, the orientational distributions of spin current in
the case of normally incident $\sigma ^{+}$ polarized light are plotted in
Figure 2.

\begin{figure}[tbp]
\includegraphics[width=7cm]{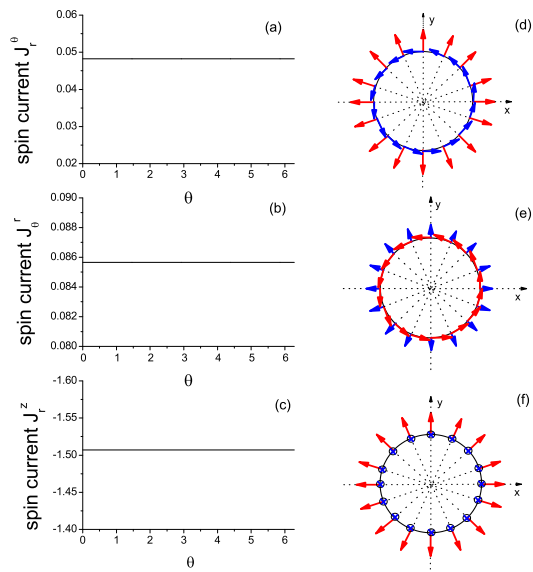}
\caption{{}Orientational distribution of pure spin current in the case of
normally incident $\protect\sigma _{+}$ polarized light. (a-c) for numerical
calculation results (in unit of $10^{-3}$eV$^{-2}$.nm$^{-1}$.fs$^{-2}$.$%
\hbar a_{0}^{2}e^{2}E_{0}^{2}\protect\tau _{e}$). The experiment parameters
are taken as: $\Delta _{0}=750$meV, $\protect\lambda =6.3$meV.nm$/\hbar $, $%
m^{\ast }=0.05m$, $\protect\gamma _{1}=6.9$, $\protect\gamma _{2}=2.2$, $%
L=14 $nm, the index of refraction $n=\protect\sqrt{13}$, and the wave length
of the light $\protect\lambda _{0}=1620$nm. (d-f) for the sketches of
orientational distribution, in which red arrows denote the directions of
spin current, blue arrows denote the polarization directions of spin
current, and $\otimes $ denotes spin polarization along $-z$ direction. (a)
and (d) for $J_{r}^{\protect\theta }\left( \protect\theta \right) $; (b) and
(e) for $J_{\protect\theta }^{r}\left( \protect\theta \right) $; (c) and (f)
for $J_{r}^{z}\left( \protect\theta \right) $. }
\end{figure}

\section{Spin linear photogalvanic effect (SLPGE)}

Now we come to consider the normal incidence of a linearly polarized light
onto the sample, and the pump pulse is of the form
\begin{equation}
\mathbf{E}\left( t\right) =\mathbf{E}_{\omega }\exp \left[ -i\left( \omega
t-kz\right) \right] +c.c.
\end{equation}%
In the medium, $E_{x}=t_{0}E_{0}\cos \phi $ and $E_{y}=t_{0}E_{0}\sin \phi $%
, with $t_{0}=2/(1+n)$ and $\phi $ is the angle between the polarization
plane and the $x$-axis, e.g. $\phi =0$ corresponding to the $x$ polarized
light. In this case it was known that no photocurrent is injected as in Eq. (%
\ref{current}). However, a PSC may survive. The physical origin of spin
current is given briefly as follows: Due to Rashba spin-orbit coupling,
conduction band splits into two subbands denoted by $\left\vert \uparrow
\right\rangle $ and $\left\vert \downarrow \right\rangle $. When the
frequency $\omega $ of the light satisfies the condition $\hbar \omega
>\Delta _{0}$, electrons are pumped from the heavy-hole band into conduction
bands. If there appears a electron state $\left\vert \mathbf{k},\uparrow
\right\rangle $ in conduction band $\left\vert \uparrow \right\rangle $ with
momentum $\mathbf{k}$, $\left\vert -\mathbf{k},\downarrow \right\rangle $
must appear in conduction band $\left\vert \downarrow \right\rangle $ with
momentum $-\mathbf{k}$ with the same probability according to the symmetry. $%
\left\vert \mathbf{k},\uparrow \right\rangle $ and $\left\vert -\mathbf{k}%
,\downarrow \right\rangle $ are two degenerate states and have opposite
velocities, thus the pair contributes a null electric current. However, they
have opposite spin polarization such that they carry equal spin current.
Therefore a finite spin current survives for these two states. The spin
splitting of conduction band plays an essential role in this mechanism.\cite%
{Cui06xxx,Li06APL} As an example, the orientational distributions of spin
current in the case of normally incident $x$ polarized light are plotted in
Figure 3.
\begin{figure}[tbp]
\includegraphics[width=7cm]{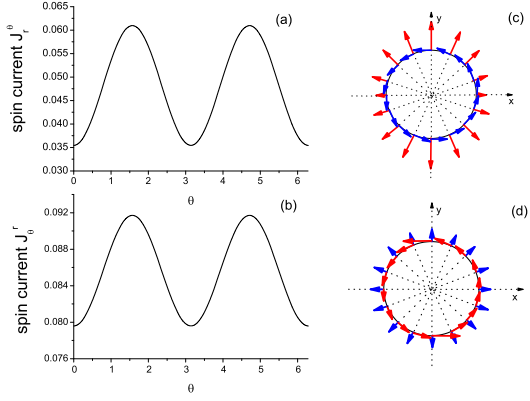}
\caption{{}Orientational distribution of pure spin current in the case of
normally incident $x$ polarized light. (a) and (b) for numerical calculation
results (in unit of $10^{-3}$eV$^{-2}$.nm$^{-1}$.fs$^{-2}$.$\hbar
a_{0}^{2}e^{2}E_{0}^{2}\protect\tau _{e}$). The experimental parameters are
taken as the same as those given in Fig. 2. (c) and (d) for the sketches of
orientational distribution, in which red arrows denote the directions of
spin current, and blue arrows denote the polarization directions of spin
current. (a) and (c) for $J_{r}^{\protect\theta }\left( \protect\theta %
\right) $; (b) and (d) for $J_{\protect\theta }^{r}\left( \protect\theta %
\right) $.}
\end{figure}

An explicit calculation gives PSC with in-plane spin polarization,
\begin{equation}
J_{x}^{y}=\left( -I_{0}^{L}-I_{1}^{L}\cos 2\phi \right) t_{0}^{2}\hbar
a_{0}^{2}e^{2}E_{0}^{2}\tau _{e},  \label{SCxy}
\end{equation}%
\begin{equation}
J_{x}^{x}=I_{1}^{L}t_{0}^{2}\hbar a_{0}^{2}e^{2}E_{0}^{2}\tau _{e}\sin 2\phi
,
\end{equation}%
where $I_{0}^{L}=\lambda \mu \left( m^{\ast }-\mu \right) /\left( 6\hbar
^{3}m^{\ast }\right) $ and $I_{1}^{L}=\lambda \mu ^{3}\Omega /\left( 6\hbar
^{5}m^{\ast }\left\langle k_{z}^{2}\right\rangle \right) $. It is obvious
that PSC with an in-plane spin polarization is dependent on the angle $\phi $
between the polarization plane and the $x$-axis. A linearly polarized light
can be decomposed as a combination of two circularly polarized beams of
light. The phase difference between these two composite beams of the light
is $2\phi $. The polarization dependence of the PSC originates from the
interference of two composite circularly polarized lights.

\section{Relation between photocurrent and spin current}

The two formulas for photocurrent in Eq. (\ref{current}) and spin current in
Eq. (\ref{SCxy}) contain the parameter $a_{0}$ and the relaxation time $\tau
_{e}$ which need to be determined experimentally. Assume the intensity and
the frequency of the two applied lights are equal, the ratio of the
photocurrent of circularly polarized light with an oblique angle $\Theta
_{0} $ to the spin current of normally incident linear polarized light gives
\begin{equation}
\frac{J_{x}}{J_{x}^{y}}=\eta \frac{t_{s}t_{p}}{t_{0}^{2}}P_{\text{circ}}\sin
\Theta \frac{2e}{\hbar },  \label{JxJxy}
\end{equation}%
where $\eta $ is a dimensionless frequency-dependent factor,
\begin{equation}
\eta =\frac{2\Omega }{\epsilon _{0}+\Omega \cos 2\phi },  \label{factor}
\end{equation}%
with $\epsilon _{0}=\hbar ^{2}\left\langle k_{z}^{2}\right\rangle \left(
m^{\ast }-\mu \right) /\mu ^{2}$. For a small incidence angle $\Theta _{0}$
and $P_{\text{circ}}=1$, the ratio is reduced to $J_{x}/J_{x}^{y}\approx
\eta \frac{\Theta _{0}}{n}\frac{2e}{\hbar }.$ In this way we establish an
explicit relation between light-injected photocurrent and PSC. All
parameters in $\eta $ are known in semiconductor materials. For the sample
of InGaAs,\cite{Yang06PRL} for instance, $\Delta _{0}=750$meV, $\lambda =6.3$%
meV.nm$/\hbar $, $m^{\ast }=0.05m$, $\gamma _{1}=6.9$, $\gamma _{2}=2.2$, $%
L=14$nm, $\epsilon _{0}=50.8$meV, and $\Omega =\hbar \omega -749.982$meV$.$
In this system, the photocurrent $J_{x}$ in CPGE has been measured
successfully. Therefore we can deduce the spin current by measuring the
photocurrent experimentally.

\section{Numerical results}

The formula in Eq. (\ref{factor}) is only valid for excitation of electrons
near the $\Gamma $ point. In principle we can calculate the ratio of
photocurrent to spin current following the $\mathbf{k}\cdot \mathbf{p}$
calculation done by\ Baht et al.\cite{Bhat05PRL} Here we present our results
after the quantum size effect of QW with a finite thickness $L$ is taken
into account. For a confining potential $V(z)$ along the $z$-axis, say $%
V(z)=+\infty \ $for $\left\vert z\right\vert $\ $>$\ $L/2$ and $V(z)=0\ $for
otherwise. While $k_{x}$ and $k_{y}$ remain to be good quantum numbers, the
quantization along the $z$-axis can be calculated numerically by the
truncation approximation if $L$ is of order of tens nm.\cite%
{Li06APL,Winkler02PRB} The lowest four valence subbands are plotted in
Figure 4, where each subband is doubly degenerated.

\begin{figure}[tbp]
\includegraphics[width=7cm]{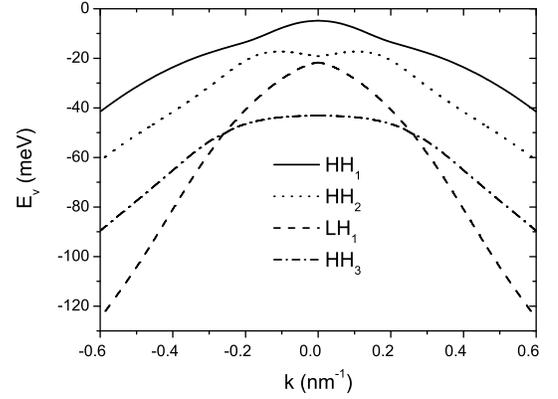}
\caption{{}Dispersion of low lying subbands for the Luttinger effective
Hamiltonian with the width $L=14$nm, $\protect\gamma _{1}=6.9$ and $\protect%
\gamma _{2}=2.2$. HH and LH denote heavy hole and light hole, respectively.}
\end{figure}

In this way photocurrent and PSC can be calculated numerically in terms of
the unknown parameters $a_{0}$ and $\tau _{e}$. The variations of $J_{x}$ in
the case of oblique incident $\sigma _{+}$ polarized light with the
frequency of light is plotted in Figure 5(a), and the photocurrent has its
sign change when the dominant contribution of interband transition to the
conduction band switches from the first heavy-hole sub-band to the second
heavy-hole sub-band. We also plot the spin current $J_{x}^{y}$ in a normally
incident $x$ polarized light in Figure 5(b). The frequency dependence of the
dimensionless factor $\eta $ is plotted in Figure 5(c). When $\hbar \omega $
is close to the band gap $\Delta _{0}$ the main contribution results from
only interband transition from the first heavy-hole subband to the
conduction band, the ratio factor is linear in the frequency $\omega $. The
photocurrent was observed experimentally in the two samples of InGaAs with
Rashba coupling $\lambda _{1}=3.0$meV.nm$/\hbar $ and $\lambda _{2}=6.3$%
meV.nm/$\hbar $.\cite{Yang06PRL} The photocurrent changes its sign when the
frequency of laser increases. The angle dependence of photocurrent gives $%
J_{x}(\Theta _{0})\simeq $ $351\Theta _{0}/n$ nA for a small angle $\Theta
_{0}$ (with the index of refraction $n=\sqrt{13}$). The ratio factor is
estimated as $\eta \simeq 0.62$. If we keep our conditions of laser except
that the helicity of light changes from circular to linear, the PSC\ in the
linearly polarized light is estimated as $J_{x}^{y}\simeq 566\hbar /2e$ nA.

\begin{figure}[tbp]
\includegraphics[width=7cm]{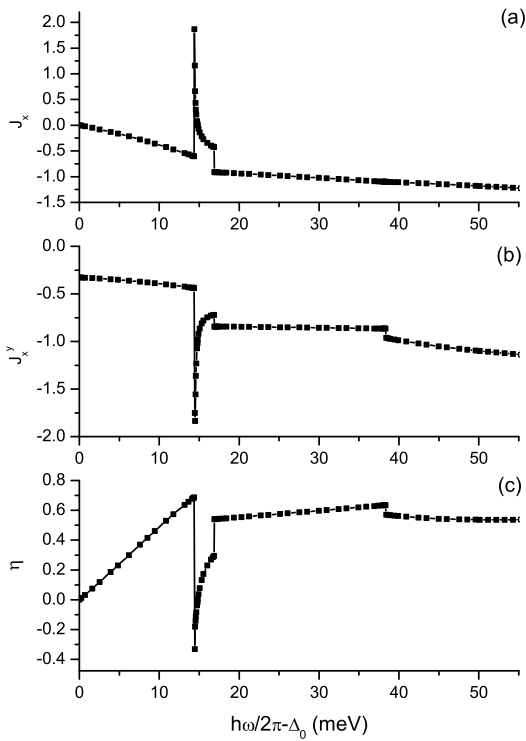}
\caption{{}Numerical results based on the model of QW with a finite
thickness $L$. The parameters are given in the text following Eq. (12). (a)
Spectrum of photocurrent $J_{x}$ in the case of oblique incident $\protect%
\sigma _{+}$ polarized light (in unit of $10^{-3}$eV$^{-2}$.nm$^{-1}$.fs$%
^{-2}$.$t_{s}t_{p}a_{0}^{2}e^{3}E_{0}^{2}\protect\tau _{e}\sin \Theta $);
(b) Spectrum of spin current $J_{x}^{y}$ in a normally incident $x$
polarized light (in unit of $10^{-3}$eV$^{-2}$.nm$^{-1}$.fs$^{-2}$.$%
t_{0}^{2}\hbar a_{0}^{2}e^{2}E_{0}^{2}\protect\tau _{e}$); (c) The frequency
dependence of the dimensionless factor $\protect\eta (\protect\omega )$. }
\end{figure}

\section{Conclusion}

Here we use the model with the twofold conduction band described by Rashba
coupling and the valence band by the Luttinger Hamiltonian to investigate
spin photogalvanic effect induced by polarized lights via interband
excitations in the semiconductor QWs. We established a relation between
light-injected photocurrent and PSC. As the photocurrent in CPGE was
extensively studied both theoretically and experimentally, we can make use
of it to deduce PSC in the same system by using the linearly polarized light
to replace the circularly polarized light in CPGE, which can be realized by
adding a 1/4-wave plate. Thus this method may provide a source of spin
current to study spin transport in semiconductors quantitatively.

\section*{Acknowledgments}

This work was supported by the Research Grant Council of Hong Kong under
Grant Nos.: HKU 7039/05P and HKU 7042/06P.


\begin{thebibliography}{99}
\bibitem{Awschalom02Book} \textit{Semiconductor Spintronics and Quantum
Computation}, edited by D. D. Awschalom, D. Loss, and N. Samarth
(Springer-Verlag, Berlin, 2002).

\bibitem{Ivchenko78SPJ} E. L. Ivchenko and G. E. Pikus, Pis'ma Zh. \'{E}ksp.
Teor. Fiz. \textbf{27}, 640 (1978) [Sov. Phys. JETP Lett. \textbf{27}, 604
(1978)].

\bibitem{Belinicher78PLA} V. I. Belinicher, Phys. Lett. A \textbf{66}, 213
(1978).

\bibitem{Averkiev83SP} N. S. Averkiev and M. I. D'yakonov, Fiz. Tekh.
Poluprov. \textbf{17}, 629 (1983) [Sov. Phys. Semicond. \textbf{17}, 393
(1983)].

\bibitem{Hagele98APL} D. H\"{a}gele, M. Oestreich, W. W. R\"{u}hle, N.
Nestle, K. Eberl, Appl. Phys. Lett. \textbf{73}, 1580 (1998).

\bibitem{Kikkawa99nature} J. M. Kikkawa and D. D. Awschalom, Nature (London)
\textbf{397}, 139 (1999).

\bibitem{Ganichev00APL} S. D. Ganichev, H. Ketterl, W. Prettl, E. L.
Ivchenko, and L. E. Vorobjev, Appl. Phys. Lett. \textbf{77}, 3146 (2000).

\bibitem{Ganichev01PRL} S. D. Ganichev, E. L. Ivchenko, S. N. Danilov, J.
Eroms, W. Wegscheider, D. Weiss, and W. Prettl, Phys. Rev. Lett. \textbf{86}%
, 4358 (2001).

\bibitem{Ganichev03JPCM} S. D. Ganichev and W. Prettl, J. Phys.: Condens.
Matter \textbf{15}, R935 (2003).

\bibitem{Yang06PRL} C. L. Yang, H. T. He, L. Ding, L. J. Cui, Y. P. Zeng, J.
N. Wang, and W. K. Ge, Phys. Rev. Lett. \textbf{96}, 186605 (2006).

\bibitem{Hubner03PRL} J. H\"{u}bner, W. W. R\"{u}hle, M. Klude, D. Hommel,
R. D. R. Bhat, J. E. Sipe, and H. M. van Driel, Phys. Rev. Lett. \textbf{90}%
, 216601 (2003).

\bibitem{Stevens03PRL} M. J. Stevens, A. L. Smirl, R. D. R. Bhat, A.
Najmaie, J. E. Sipe, and H. M. van Driel, Phys. Rev. Lett. \textbf{90},
136603 (2003).

\bibitem{Najmaie05PRL} A. Najmaie, E. Ya. Sherman, and J. E. Sipe, Phys.
Rev. Lett. \textbf{95}, 056601 (2005).

\bibitem{Bhat05PRL} R. D. R. Bhat, F. Nastos, A. Najmaie, and J. E. Sipe,
Phys. Rev. Lett. \textbf{94}, 096603 (2005).

\bibitem{Zhao05PRB} H. Zhao, X. Pan, A.L. Smirl, R.D.R. Bhat, A. Najmaie,
J.E. Sipe, and H.M. van Driel, Phys. Rev. B \textbf{72}, 201302(R) (2005).

\bibitem{Tarasenko05JETP} S. A. Tarasenko and E. L. Ivchenko, JETP Lett.
\textbf{81}, 231 (2005).

\bibitem{Tarasenko06xxx} S. A. Tarasenko and E. L. Ivchenko,
cond-mat/0609090.

\bibitem{Kato04Science} Y. K. Kato, R. C. Myers, A. C. Gossard, and D. D.
Awschalom, Science \textbf{306}, 1910 (2004).

\bibitem{Wunderlich05PRL} J. Wunderlich, B. Kaestner, J. Sinova, and T.
Jungwirth, Phys. Rev. Lett. \textbf{94}, 047204 (2005).

\bibitem{Tinkham06Nature} S. O. Valenzuela and M. Tinkham, Nature (London)
\textbf{442}, 176 (2006).

\bibitem{Cui06xxx} X. D. Cui, S. Q. Shen, J. Li, W. K. Ge, and F. C. Zhang,
cond-mat/0608546.

\bibitem{Li06APL} J. Li, X. Dai, S. Q. Shen, and F. C. Zhang, Appl. Phys.
Lett. \textbf{88}, 162105 (2006).

\bibitem{Winkler2003} R. Winkler, \textit{Spin-Orbit Coupling Effects in
Two-Dimensional Electron and Hole Systems}, (Springer-Verlag, Berlin, 2003).

\bibitem{HaugBook} H. Haug and S. W. Koch, \textit{Quantum Theory of the
Optical and Electronic Properties of Semiconductors} (World Scientific,
Singapore, 1993).

\bibitem{Rossler02pss} U. R\"{o}ssler, Phys. Status Solidi B \textbf{234},
385 (2002).

\bibitem{Golub03PRB} L. E. Golub, Phys. Rev. B \textbf{67}, 235320 (2003).

\bibitem{Ivchenko97Book} E. L. Ivchenko and G. E. Pikus, \textit{%
Superlattices and Other Heterostructures. Symmetry and Optical Phenomena}
(Springer, Berlin, 1997).

\bibitem{Hache98IEEE} A. Hach\'{e}, J. E. Sipe, and H. M. van Driel, IEEE J.
Quantum Electron. \textbf{34}, 1144 (1998).

\bibitem{Hankiewicz05PRB} E. M. Hankiewicz, J. Li, T. Jungwirth, Q. Niu, S.
Q. Shen, and J. Sinova, Phys. Rev. B \textbf{72}, 155305 (2005).

\bibitem{Winkler02PRB} R. Winkler, H. Noh, E. Tutuc, and M. Shayegan, Phys.
Rev. B \textbf{65}, 155303 (2002).
\end{thebibliography}
\end{document}